\documentclass[twocolumn,showpacs,preprintnumbers,amsmath,amssymb]{revtex4}

\usepackage{graphicx}
\usepackage{dcolumn}
\usepackage{bm}

\begin{document}

\newif\ifplot
\plottrue
\newcommand{\RR}[1]{[#1]}
\newcommand{\intsum}{\sum \kern -15pt \int}
\newfont{\Yfont}{cmti10 scaled 2074}
\newcommand{\Y}{\hbox{{\Yfont y}\phantom.}}
\def\O{{\cal O}}
\newcommand{\bra}[1]{\left< #1 \right| }
\newcommand{\braa}[1]{\left. \left< #1 \right| \right| }
\def\Bra#1#2{{\mbox{\vphantom{$\left< #2 \right|$}}}_{#1}
\kern -2.5pt \left< #2 \right| }
\def\Braa#1#2{{\mbox{\vphantom{$\left< #2 \right|$}}}_{#1}
\kern -2.5pt \left. \left< #2 \right| \right| }
\newcommand{\ket}[1]{\left| #1 \right> }
\newcommand{\kett}[1]{\left| \left| #1 \right> \right.}
\newcommand{\scal}[2]{\left< #1 \left| \mbox{\vphantom{$\left< #1 #2 \right|$}}
\right. #2 \right> }
\def\Scal#1#2#3{{\mbox{\vphantom{$\left<#2#3\right|$}}}_{#1}
{\left< #2 \left| \mbox{\vphantom{$\left<#2#3\right|$}}
\right. #3 \right> }}

\title{
Resolving the Discrepancy of  135 MeV 
{\mbox{\boldmath $pd$}} 
Elastic Scattering Cross Sections and 
Relativistic Effects}

\author{K.\ Sekiguchi$^{1}$}
\author{H.\ Sakai$^{1,2,3}$}
\author{H.\ Wita{\l}a$^{2,4}$}
\author{W.\ Gl\"ockle$^{5}$}
\author{J.\ Golak$^{4}$}
\author{K.\ Hatanaka$^{6}$}
\author{M.\ Hatano$^{2}$}
\author{K.\ Itoh$^{7}$}
\author{H.\ Kamada$^{8}$}
\author{H.\ Kuboki$^{2}$}
\author{Y.\ Maeda$^{3}$}
\author{A.\ Nogga$^{9}$}
\author{H.\ Okamura$^{10}$}
\author{T.\ Saito$^{2}$}
\author{N.\ Sakamoto$^{1}$}
\author{Y.\ Sakemi$^{6}$}
\author{M.\ Sasano$^{2}$}
\author{Y.\ Shimizu$^{6}$}
\author{K.\ Suda$^{3}$}
\author{A.\ Tamii$^{6}$}
\author{T.\ Uesaka$^{3}$}
\author{T.\ Wakasa$^{11}$}
\author{K.\ Yako$^{2}$}
\affiliation{
$^{1}$RIKEN, the Institute of Physical and Chemical Research,
Wako, Saitama 351-0198, Japan}
\affiliation{
$^{2}$Department of Physics, University of Tokyo, Bunkyo,
Tokyo 113-0033, Japan}
\affiliation{
$^{3}$Center for Nuclear Study, University of Tokyo, Bunkyo,
Tokyo 113-0033, Japan}
\affiliation{
$^{4}$Institute of Physics, Jagiellonian University,
PL-30059 Cracow, Poland}
\affiliation{
$^{5}$Institut f\"ur theoretische Physik II,
Ruhr-Universit\"at Bochum, D-44780 Bochum, Germany}
\affiliation{$^{6}$Research Center for Nuclear Physics, Osaka
University, Ibaraki, Osaka 567-0047, Japan}
\address{$^{7}$Department of Physics, Saitama University, 
Saitama 338-8570, Japan}%
\affiliation{
$^{8}$Department of Physics, 
Faculty of Engineering, 
Kyushu Institute of Technology, Kitakyushu 804-8550, Japan}
\affiliation{$^{9}$Institut f\"ur Kernphysik,
Forschungszentrum, J\"ulich, D-52425 J\"ulich, Germany}
\affiliation{$^{10}$Cyclotron and Radioisotope Center, Tohoku
University, Sendai, Miyagi 980-8578, Japan}
\affiliation{$^{11}$Department of Physics, Kyushu University,
Fukuoka 812-8581, Japan}

\date{\today}

\begin{abstract}
 Three precise measurements 
 for elastic $pd$ scattering at 135 MeV/A
 have been performed 
 with the three different experimental setups.
 The cross sections are described well by the theoretical 
 predictions based on modern 
 nucleon-nucleon forces combined with three nucleon forces. 
 Relativistic Faddeev calculations show that relativistic effects 
 are restricted to backward angles.
This result supports the two measurements recently reported by RIKEN and
contradicts the KVI data.

\end{abstract}

\maketitle


A hot topic of present day few-nucleon system studies is 
to explore the properties
of three-nucleon forces (3NFs) acting in systems with more than A=2
nucleons. 
These forces appear the first time in the three--nucleon (3N) system
where they provide an additional contribution to a predominantly
pairwise potential energy of three nucleons.
Meson-exchange picture can undoubtedly lead to such forces,
however they are relatively weak compared to nucleon-nucleon (NN) 
forces, and therefore it is hard to approach and also to find evidences 
for them experimentally. 
The first evidence of the 3NF is found
in the 3N bound systems, $^3\rm H$ and $^3 \rm He$.
The binding energies of these nuclei are not described by 
exact solutions of the three--nucleon Faddeev equations employing modern NN 
forces, e.g. AV18~\cite{AV18}, CDBonn~\cite{cdb}, Nijmegen I, II 
and 93~\cite{nijm}\@(see e.g. Ref.~\cite{underbind}).  
The discrepancy between data and theory
is explained by adding 3NF, mostly based 
on $2\pi$-exchange between three nucleons with 
the $\Delta$-isobar excitation, 
such as the Tucson-Melbourne (TM)~\cite{TM} and 
the Urbana IX 3NF~\cite{uIX}.
The binding energies show the significant contributions 
of the 3NF, however
they could only constrain the overall strength. 
In order to study
the momentum and/or spin dependence of the 3NF, 
the 3N scattering is one attractive 
approach; others are to study the spectra of 
nuclear systems up to A=10~\cite{piep2002}.
%

Indication of 3NF for the 3N scattering
was first pointed out in the cross section
minima for nucleon--deuteron $(Nd)$ elastic scattering 
at energies of the incoming nucleon above $\approx 60~\rm
MeV$ by Witala {\it et al.} in 1998~\cite{wit98}\@.
Since then experimental 
studies of higher--energy proton--deuteron ($pd$) 
elastic scattering covering  
incident energies of up to $\approx 250$~MeV have been performed
extensively and provided precise data of 
cross sections~\cite{sak96,sak2000,sek2002,hatanaka02,erm2003} 
and spin observables, such as 
analyzing powers~\cite{sak96,sak2000,sek2002,anpow}, 
spin correlation coefficients~\cite{spincor}, 
and polarization transfer coefficients~\cite{hatanaka02,sek2004}.
Precise cross section data for the elastic $dp$ scattering 
taken at RIKEN with 135 MeV/A deuterons have shown 
large disagreement between data and rigorous
Faddeev calculations with modern NN forces~\cite{sak96,sak2000,sek2002}.
Combination of these NN forces and 3NFs such as 
the TM99~\cite{tm99}  (TM99 is a version of the TM force which is
more consistent with chiral symmetry~\cite{fri99,huba1}) 
and the Urbana IX removed this discrepancy 
and led to a good description of the measured cross sections.
This result can be taken as a clear signature
of the 3NF effects in $Nd$ elastic scattering. 
However spin observables are not always explained 
by addition of the 3NFs, showing
the defects in spin parts of the 3NFs~\cite{sek2004}.

The recent measurement of the elastic 
$pd$ scattering with a 135 MeV proton beam and a mixed 
solid $\rm CD_2$--$\rm CH_2$ target 
at KVI Groningen~\cite{erm2003} provided cross sections which were in
disagreement with the data at RIKEN. The KVI data were larger than
the RIKEN data by about $10$--$40\%$ and also differed in shape. 
%
%
{\bf IF} the KVI data were correct, the presently available 2$\pi$ exchange
3NFs would be insufficient to explain the difference.
Therefore one should look for other sources 
which have not been considered to fill the difference,
such as 3NFs other than 2$\pi$ exchange types and/or 
relativistic effects or something completely new.
According to the recent theoretical predictions 
in the framework of the coupled channel approach~\cite{arnas2003},
the $\pi$--$\rho$ and $\rho$--$\rho$ type 3NFs cause little effect.
Therefore relativistic effects could be a candidate to fill 
the difference.
We estimate for the first time 
the magnitude of relativistic effects 
in the 3N scattering.
Of highest importance, however, is to clarify by experiment
which are correct, the RIKEN data or the KVI data.

%
Aiming to resolve the discrepancy by experiment,
we performed the following three measurements for elastic $pd$ scattering.
First, we made a measurement at RIKEN 
with the proton beam  
and a $\rm CD_2$--$\rm CH_2$ sandwiched solid target
at the angles where the $pp$ and $pd$ elastic scattering were 
simultaneously measured with the magnetic spectrograph
SMART\@. 
Using the well-known elastic $pp$ cross sections
we can estimate the overall systematic uncertainty
for the $pd$ cross section.
We used the H$^+_2$ ions of 270 MeV as the 135 MeV proton 
beam for convenience of acceleration.
%
Secondly, to confirm the angular distribution 
we measured again with 135 MeV/A deuterons.
This measurement was performed just after the 
previous $pp$ scattering experiment with the same experimental setup
in order to minimize the systematic uncertainties.
Note the mass of $\rm H_2^+$ is almost identical to that of deuteron
so that we did not need to change any parameters of the accelerators
or beam transport system.
We tried to check the fluctuations of the 
target thickness during the experiment 
by measuring the $dp$ scattering 
at the fixed angle $\theta_{\rm c.m.} = 69.7^\circ$,
where the scattered deuterons and recoil protons 
were detected in coincidence in the scattering chamber.
For the same purpose, 
the cross section at $\theta_{\rm c.m.}= 165.1^\circ$ 
was measured with the SMART system over several times
during the experiment.
We also measured the carbon background events 
which were not obtained 
in the previous measurement~\cite{sak2000,sek2002}.
Lastly,
we performed a totally independent measurement 
at the Research Center for Nuclear Physics (RCNP) of Osaka University, 
using a 135 MeV proton beam and deuterated polyethylene target.
The absolute normalization of the cross sections has been 
performed by taking data with a $\rm D_2$ gas target and 
the double slit system for which 
the RCNP group has already established the procedure to obtain
the absolute $pd$ cross section~\cite{hatanaka02}.
In this paper we would like
to present these new data and compare them to the old ones and to
theoretical predictions including relativistic effects
to clarify firmly the 3NF effects.


The measurement of the cross sections with 270 MeV $\rm H^{2+}$ beam
($p$ on $\rm ^2H$)
was carried out at the RIKEN accelerator research facility
with the SMART system.
The target was a sandwich of the polyethylene ($\rm CH_2$)
with thickness of $\rm 18.7~mg/cm^2$ 
and the self-supporting 99\% isotopically enriched 
deuterated polyethylene foil ($\rm CD_2$)
with thickness of $21 \rm mg/cm^2$~\cite{hatanaka02,yukie02}\@.
The relative deviation of the $\rm CD_2$ target thickness 
was estimated to be within about $2.5\%$
which was attributed to the inhomogeneity of the $\rm CD_2$ foil.
At the lab. angles $10^\circ$--$14^\circ$ 
which correspond to the c.m. angles $16^\circ$--$20^\circ$ 
for the elastic $pd$ scattering 
the momentum difference of the scattered protons 
from the $pp$ and $pd$ elastic scattering
is within 4\%. In this angular range  the protons emitted 
from each of these reactions 
were measured simultaneously with the SMART spectrograph and 
their energy spectra were completely resolved.
The yields for the elastic $pd$ scattering were obtained by
subtracting carbon contributions in the excitation energy spectra.
The elastic $pp$ scattering yields were obtained 
by subtracting the backgrounds 
from the $p+d \to p+p+n$ breakup reaction 
and from the proton scattering on the carbon.
For interpolation purposes the breakup 
background contribution was assumed to be 
a third--order polynomial in the proton energy.
The measured cross sections for $pp$ elastic scattering
were compared with the values calculated by the phase-shift
analysis code SAID~\cite{said}
and found to be  consistent within 2\%.
Thus we estimated the overall systematic uncertainty 
of the measured cross section data to be 3\% at most.

We made the cross section measurement
with  270 MeV deuteron beam ($d$ on $\rm ^1H$) 
in the angular range $\theta_{\rm c.m.} = 10^\circ$--$180^\circ$.
The $\rm CH_2$ target used in the proton beam 
experiment was employed as a hydrogen target.
For the forward scattering ($\theta_{\rm c.m.} \le 90^\circ$)
the scattered deuterons were
detected while for the backward scattering 
($\theta_{\rm c.m.} \ge 90^\circ$)
the recoil protons were measured. 
The statistical errors of the cross sections are within 1.6\%.
The fluctuation of the target thickness is within 3\%.
The uncertainty due to carbon background subtraction is less than 5\%. 
The overall systematic uncertainties which include also
the uncertainties by $pp$ experiment are estimated to be 6\%.
%

The experiment performed at RCNP used a proton beam 
in conjunction with the high resolution spectrometer 
Grand Raiden\@. 
The proton beam was accelerated up to 135 MeV
by the AVF and ring cyclotrons
and bombarded the same $\rm CD_2$ foil 
we used in the experiment at RIKEN. 
The proton beam was stopped in a Faraday cup in the 
scattering chamber, except for the gaseous target 
measurement at the angle $\theta_{\rm lab.} = 25.5^\circ$.
In this case the beam was stopped in a Faraday cup 
located downstream outside the scattering chamber.
The scattered protons
or deuterons were momentum analyzed by the Grand Raiden.
The protons were measured at the angles 
$\theta_{\rm c.m.} \le 90^\circ$ and
the recoil deuterons were detected at 
the angles $\theta_{\rm c.m.} \ge 90^\circ$.
The measured angles were 
$\theta_{\rm c.m.} = 17.0^\circ$--$157.7^\circ$.
The yields from $\rm D_2$ were obtained by 
subtracting carbon contributions in the excitation energy spectra.
%
To normalize cross sections taken with the $\rm CD_2$ target
a measurement with a $\rm D_2$ gas target was performed 
at the lab. angles $25.5^\circ$ and $60^\circ$
which corresponded to the c.m. angles $39.0^\circ$ and $87.4^\circ$, 
respectively.
The $\rm D_2$ gas target was contained in the cell
of a cylinder of $40~\rm mm$
diameter made of $\rm 200~\mu m$--thick aluminum.
The absolute gas pressure was continuously monitored
with a precision better than 0.2\%\@.
The temperature of the target cell was checked during the 
measurement and kept at room temperature.
A double--slit system was used to determine precisely the target
volume and the solid angle of the Grand Raiden spectrometer.
The effective target thickness and the solid angle were
calculated by Monte Carlo simulations.
Spectra with the empty gaseous cell were also measured to
determine background contributions from the aluminum cell.
An additional measurement was performed with hydrogen
gas replacing the deuteron one in order to cross--check 
the experimental setup at the angle $\theta_{\rm lab.} = 25.5^\circ$. 
The measured cross section of $pp$ scattering 
is consistent  within 3\% 
with the calculated results by the SAID program.
The statistical errors of the $pd$ elastic scattering
cross sections are smaller than  $1.4\%$.
The absolute normalization was estimated to be 3\% 
by elastic $pp$ scattering measurement.
The uncertainty  due to carbon background subtraction 
for the excitation energy spectrum is 3\%.
There is also the uncertainty of 2.5\% 
attributed to the inhomogeneity of the $\rm CD_2$ foil.
The overall systematic uncertainties are estimated to be 
5\% at most.


All the  experimental results are shown in Fig.~\ref{dcs_exp}.
The data taken at RCNP are shown with solid diamonds. 
The open squares (circles) are the data measured with the proton 
(deuteron) beam at RIKEN.
The data published 
in Refs.~\cite{sak96,sak2000,sek2002}
are shown with open diamonds.
The KVI data reported 
in Ref.~\cite{erm2003} are shown with 
open triangles.
Only statistical errors are presented.
The very good agreement between the independent measurements 
allows us to conclude that  
 the systematic uncertainty due to the detection
setup is small.
Comparison of the new three  sets of data 
to the previously reported ones~\cite{sak96,sak2000,sek2002} 
supports the previous measurements and shows a clear disagreement 
with the KVI data~\cite{erm2003}.

In Fig.~\ref{dcs_exp} we compare the data to theoretical predictions 
based on modern NN forces and their combinations with 
3NFs
\cite{wit88,hub97,hub93,glo96}. 
We used the modern NN forces AV18,
CDBonn, Nijmegen I and II combined with the 3NF TM99 ~\cite{tm99} 
with the cut-off $\Lambda$ values which 
lead for a particular NN force combined with the 3NF TM99  
to a reproduction of the  $^3\rm H$ binding 
energy. In case of the AV18 NN force
we also combined it with the 3NF Urbana IX . 

As is seen in  Fig.~\ref{dcs_exp}   
various NN force
predictions (light shaded band) clearly underestimate the cross section data
for the angles $\theta_{\rm c.m.} \ge 90^\circ$. 
The narrow band of predictions
reflects the weak dependence on the choice of the nearly on-shell equivalent
NN interaction. The inclusion of the TM99 (the dark shaded band)
or, in case of AV18, of the 3NF  Urbana IX  (solid line), leads to a very good
description of the data. 

In view of such a good description of the cross section it is interesting to
find out how large relativistic effects are at 135 MeV/A. 
This was done assuming
that only NN forces are acting. We followed a formalism for treating the
relativistic three-body Faddeev equations of Ref.~\cite{relform} with a
boosted two-nucleon potential $V$ expressed in terms of the relativistic
potential $v$ given in the NN c.m. system, 
\begin{eqnarray}
V(\overrightarrow P) &\equiv& \sqrt{[\omega (\overrightarrow k) + v]^2 + 
{\overrightarrow P}^2} 
 - \sqrt{ {\omega(\overrightarrow k)}^2 + {\overrightarrow P}^2} .~
\label{eq1}
\end{eqnarray}
The momentum $\overrightarrow P$ is the total momentum of the
two-nucleon system, and $\overrightarrow k$ and $-\overrightarrow k$ are the
individual momenta of the nucleons in their NN c.m.
($\omega (\overrightarrow k) = 2 \sqrt{\overrightarrow k^2 + m^2}$). 
We did not calculate
the matrix elements of the boosted potential in all its
complexity~\cite{kam2002} but restricted only to the leading order terms 
in a $P/\omega $ and $v/\omega$ expansion,
\begin{eqnarray}
V(\overrightarrow k, \overrightarrow{k^\prime}; \overrightarrow P)  
&=&  v(\overrightarrow k, \overrightarrow{k^\prime})
\biggl\{ 1 - 
{ \overrightarrow P ^2\over{2 \omega(\overrightarrow k)  
\omega(\overrightarrow {k^\prime}) } } \biggr\} .
\label{eq2}
\end{eqnarray}
%
We checked  the quality of the approximation of Eq.(\ref{eq2}) by calculating
the deuteron wave function $\phi_d(\overrightarrow k)$ when the deuteron
is moving with  a momentum corresponding to 135 MeV/A. 
The resulting deuteron binding energy and the deuteron $D$-state 
probability for the deuteron in such a motion are close to the values 
for the deuteron at rest.
Choosing the exact expression Eq.(1) those properties come out exactly.
A relativistic potential $v$ was generated from the nonrelativistic 
NN potential AV18  by performing the scale transformation
of Ref.~\cite{kam98}.
%
This should be improved in future studies by allowing for
dynamically dictated relativistic features in the potential. 
Using $V(\overrightarrow k, \overrightarrow{k^\prime}; \overrightarrow P)$ 
the boosted $t$-matrix elements 
$t(\overrightarrow k, \overrightarrow{k^\prime}; \overrightarrow P)$ are 
calculated and they form the dynamical input for the 3N 
Faddeev equation with the relativistic form of the free propagator
$G_0$~\cite{kamfewb}.  
To solve this equation in the  relativistic case it is  most convenient to
use instead of the  standard Jacobi momenta~\cite{wit88} 
the relative momentum $\overrightarrow k$ in the NN
c.m. subsystem  and the 
momentum $\overrightarrow q$  of the spectator nucleon in the 3N
c.m. system. 
In the nonrelativistic limit the momentum $\overrightarrow k$ 
  reduces to the standard Jacobi momentum  
$\overrightarrow p$~\cite{glo96}.
The relativistic formulation applied is of the  Bakamjian-Thomas
type and belongs to the instant form of
relativistic dynamics~\cite{keister91}.

Nowadays partial wave decomposition is still required to 
solve numerically 3N Faddeev equations.
The standard  partial wave states~\cite{glo96}, however, are
generalized due to the choice of the NN--subsystem momentum 
$\overrightarrow k$ and
the total spin $s$ both defined in the NN c.m. system. This lead to Wigner
spin rotations when boosting to the 3N
c.m. system~\cite{keister91,wit04}, resulting in
a more  complex form for the permutation matrix element~\cite{wit04}
than used in~\cite{glo96}. 
The details of our relativistic formulation and its numerical
performance are given in Ref.~\cite{wit04}.
%
%


A restricted relativistic calculation with $j < 2$ partial 
waves states showed that Wigner spin rotations have
only negligible effects on the cross section at 135 MeV/A.
Thus when performing the fully converged calculation
($j \le 5, J \le 25/2$) we neglected the  Wigner rotations completely. The
resulting cross sections are shown in Fig.\ref{dcs_exp}. 
It is seen that the effects of
relativity are visible only in the backward angular region for
$\theta_{\rm c.m.} \ge 160^\circ$ where they increase the cross section by
up to about $15\%$. 
For $\theta_{\rm c.m.} < 160^\circ$  the effects of relativity
are practically negligible.


Summarizing, we performed measurements 
of the cross sections for elastic $pd$
scattering using 135 MeV proton and 270 MeV deuteron beams.
Present experimental arrangements allowed us to get precise 
cross section data. 
The three sets of data presented here taken with three 
different experimental setups
completely 
support the previous measurement and disagree with 
the KVI data.
The agreement of our new sets and our old set,
measured with different experimental setups, gives
confidence that the systematic errors are small.

Comparison of our data with theory based on different NN forces
combined with current 3NFs revealed clear evidence for the action of
3NFs. The discrepancies between the  135 MeV cross section data and the pure
two-nucleon force predictions can be removed by including the 3NFs 
TM99 or Urbana IX\@. 
The conclusion that 3NF effects are seen in the region 
of the cross section minimum is %
further supported by including relativity in the instant
form of relativistic dynamics as proposed by Bakamjian and Thomas.
This leads to small relativistic effects at backward angles,
but negligible contributions in the minimum, leaving 
3NFs as the only plausible mechanism to resolve the 
discrepancies between NN theory and data.
Our results clearly indicate
the usefulness of $Nd$ elastic scattering for the  study
of 3NFs. 
$Nd$ elastic scattering cross sections together with spin observables
at higher energies, where there are still discrepancies,  will
provide  an important information
to test forthcoming additional 3N force mechanisms.

This work was supported financially in part by 
the Grants-in-Aid 
for Scientific Research Numbers 04402004 and 10304018 of 
the Ministry of Education, Culture, Sports, Science, and Technology of Japan, 
and by the Polish Committee for Scientific Research under
Grant No.\ 2P03B00825, 
and by US-DOE grants No.\ DE-FC02-01ER41187 and
DE-FG02-00ER41132. 
The numerical calculations have been performed on the 
CRAY SV1 and the CRAY T3E of the NIC in J\"ulich,
Germany.  


\begingroup
\squeezetable
\endgroup

\bibliography{apssamp}

\subsection*{FIGURES}

\begin{figure}[htbp]
\caption{
(Color online)
Differential cross section for elastic $Nd$ scattering at
135 MeV/nucleon.
The light shaded band contains NN force
predictions (AV18, CDBonn, Nijmegen I and II). The dark shaded band 
results when they are combined with the TM99 3NF. 
The solid line is the AV18 + Urbana IX 
prediction. 
The dashed and dotted lines are the results of 
relativistic and nonrelativistic Faddeev calculations based on the NN forces AV18.
The  symbols are data from different measurements 
(see text).
}
\label{dcs_exp}
\end{figure}

\end{document}